# Ultrasensitive charge detection utilizing coupled nonlinear micromechanical resonators

Xuefeng Wang[1], Dong Pu[1], Ronghua Huan[1*], Xueyong Wei[2†], Weiqiu Zhu[1]

1 Department of Mechanics, Key Laboratory of Soft Machines and Smart Devices of Zhejiang Province, Zhejiang University, Hangzhou 310027, People's Republic of China

2 State Key Laboratory for Manufacturing Systems Engineering, Xi'an Jiaotong University, Xi'an 710049, People's Republic of China

**Corresponding author**: *rhhuan@zju.edu.cn,  † seanwei@mail.xjtu.edu.cn

**Abstract:**

Since the discovery of electrons, an accurate detection of electrical charges has been the dream of scientific community. Due to some remarkable advantages, micro/nano-electromechanical systems (M/NEMS) based resonators have been used to design electrometers with exquisite sensitivity and resolution. Inevitably, some limits including requisite ultra-low environmental temperature, complicated resonator structure and measurement circuit, required linear dynamic response, will cause a gap with respect to practical application. Here, we demonstrate a novel ultra-sensitive charge detection based on the linear dependence of peak frequency drift on the coupling voltage variation of two coupled nonlinear micromechanical resonators. We achieved ultra-high resolution of 2.051E-3 $fC$ (about 13 electrons) of charge detection at the room temperature. We also show an extra application of this device for weak and ultra-low-frequency (ULF) signal detection. Our findings provide a simple strategy for measuring electron charges in an extreme accuracy and developing electrometers in smaller sizes.

Highly sensitive electrometers have been researched for more than one century and they are prevalent in many diverse applications, such as mass spectrometry, surface charge analysis, particle detection of nuclear studies, and various applications of powder technology and aerosol science[1]. Due to the advantages of low cost, fast response, high accuracy and batch manufacturing[2], micro/nano-electromechanical systems (M/NEMS) based resonators have been used to design electrometers with exquisite sensitivity and resolution[3]. Inevitably, some limits including requisite ultra-low environmental temperature[3,4], complicated resonator structure and measurement circuit[5], required linear dynamic response[6] or unachievable real-time detection[7], will cause a gap with respect to practical application. Therefore, a simple, real-time, high-resolution electrometer, which can be used at the room temperature, is useful for practical application. Besides, a fact that most resonators are desired to work in linear regime up to now also deserves a lot of attention to extend. Due to size effect[8], micro-resonators are easier to be excited into nonlinear regime[9]. What's more, exploitation of nonlinear phenomena[10,11,12] to improve performance has recently received significant attention, including mass sensing in terms of coupled nonlinear MEMS resonators[13], novel signal amplification scheme through bifurcation topology[14], etc. Obviously, the bifurcation exists widely in nonlinear system[15,16], and it is worthwhile to excavate the potential of nonlinear phenomena for practical applications. Here, we show that a coupled micro-resonators system can greatly enhance the resolution of an electrometer and realize real-time measurements. Its resolution is improved by nearly 3 orders of magnitude compared to counterparts reported in the earlier literatures. Meanwhile, the proposed one's structure is simpler and it can be operated in the nonlinear region.

**Results and Discussion**

*The Basic Mechanism of High-resolution Charge Detection*

The open-loop measurement circuit shown in Fig.1a is built to characterize this coupled system. The basic structure of the electrometer consists of two identical silicon-based micro-resonators (see in the Supplement Material S1). An electric voltage $V_{DC} + V_{AC}cos(\omega t)$ is loaded on one electrode to drive the resonator R1 to vibration. The transverse displacement of resonator beam is sensed due to the piezoresistive effect. A DC voltage $V_C$ is loaded on the body of the other resonator R2 to control the coupling strength. In Fig.1b, we plot the peak frequency of resonant amplitude curves for increasing coupling voltage $V_C$. A 1895 Hz discontinuous jump of the peak frequency is observed when $V_C$ reaches a threshold. The inset of Fig.1b further demonstrates this discontinuous phenomenon for $V_C$ below and above the threshold.

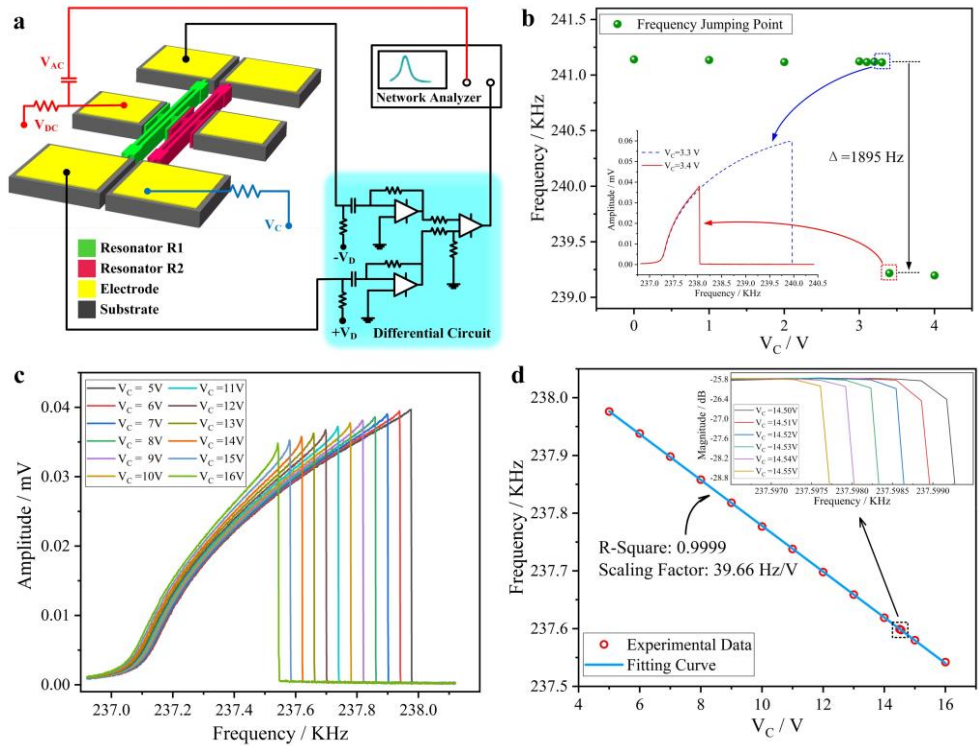

**Fig.1 Open-loop experiment.** All tests are processed with the same conditions of 316.2 mV $V_{AC}$ (driving voltage) and 18 V $V_{DC}$ (polarization voltage). **a** Schematic drawing of the open-loop measurement circuit. **b** Peak frequency for different coupling voltage. The inset shows the resonant amplitude curves for the coupling voltage below the threshold (blue dash line) and above the threshold (red solid line). **c** Measured responses of resonator R1 for varying $V_C$ from 5 V to 16 V with 1 V step length. **d** Measured peak frequency shift vs. the coupling voltage.

Set $V_C$ above the threshold. Resonant amplitude curves for uniformly increasing $V_C$ with a step of 1 V are measured and plotted in Fig.1c. With the increasing of $V_C$ the resonant frequency is continuously decreased. Fig.1d plotted the peak frequency as a function of $V_C$, which reveals extremely high linear relation between the peak frequency drift and $V_C$ with fitting R-square coefficient 0.9999. The inset in Fig.1D shows the different resonant amplitude curves when the coupling voltage varies with a very small step 0.01 V (the corresponding variation of charge ΔQ is 104.6E-3 fC according to $\varDelta Q=C\varDelta V_C$, where $C$=0.01047pF is capacitance between two sensing electrodes (Supplement Material S1)), where linear variation can still be detected clearly. This linear phenomenon can be used for charge detection. The resolution $R$ can be determined by

$$R = C \cdot \delta f / K_{sf} \qquad (1)$$

where $\delta f$ is frequency fluctuation of resonator R1; $K_{sf}$ represents the scaling factor of peak frequency shift to coupling voltage $V_C$, which is equal to the absolute value of the slope of linear line in Fig.1d. In this letter, $K_{sf}$ is calculated as 39.66 Hz/V. In open-loop experiment, $\delta f$ equals to standard deviation of peak frequency extracted from 20-minute continual frequency sweep. With an output $\delta f$ of 158 mHz, the resolution is then calculated to be 35.9E-3 fC, which is much better than the counterparts reported in the earlier literatures[5,6,7].

A non-dimensional dynamic model of the coupled nonlinear micro-resonators is established [17]

$$\ddot{x} + \frac{1}{Q}\dot{x} + x + \gamma_1 x^3 = F\cos(\omega t) + J_c(x - y)$$
$$\ddot{y} + \frac{1}{Q}\dot{y} + p^2 y + \gamma_2 y^3 = J_c(y - x) \qquad (2)$$

where $x$, $y$ are the non-dimensional displacements of resonator R1 and resonator R2, respectively; $Q$ is quality factor; $p = \omega_2/\omega_1 > 1$, $\omega_1$, $\omega_2$ are the natural resonant frequencies of resonator R1 and resonator R2, respectively; $\gamma_1$, $\gamma_2$ are the normalized cubic nonlinear stiffness coefficients; $F$ is the normalized amplitude of the electrostatic excitation; $\omega$ is the normalized electrostatic excitation frequency; $J_c$ is the strength of electrostatic coupling.

Numerical simulations of resonant amplitude curves of this system are obtained and plotted (Fig.2) through numerical method[18,19] with parameters extracted from experiment. A bifurcation point $P_1$ is observed when the coupling voltage $V_C$ reaches a threshold, which can exactly explain the discontinuous jump phenomenon of the peak frequency in Fig 1b. The bifurcation point is due to the energy transmission between these two resonators. With increasing coupling strength, more energy is transferred from driving resonator to another one. When the coupling voltage reaches a threshold, a large amount of vibration energy is transferred to the resonator R2. At this moment, the energy obtained by the resonator R1 is not enough to produce large amplitude, which leads to the discontinuous phenomenon[20]. The inset a of Fig.2 shows the frequency of bifurcation point decreases linearly as the coupling voltage increases with fitting R-square coefficient 0.997, which agrees with the experiment data (Fig.1D).

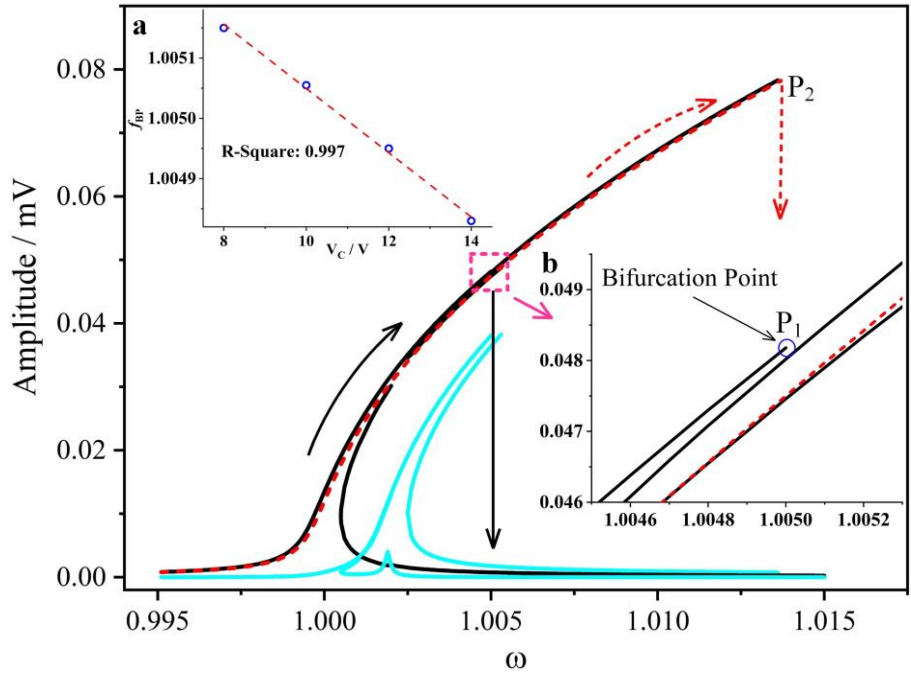

**Fig.2 Numerical resonant amplitude curves of both resonators.** Solid black lines represent the response of resonator R1. Solid cyan lines represent the response of resonator R2. Dash red line represent the partial response of resonator R1 with coupling voltage under the threshold. These arrows show the trend of responses. **a** This inset shows the frequency on bifurcation point decreases linearly as the coupling voltage increases. These circles represent results from numerical simulation. The red solid is a fitting curve. **b** This inset is an enlarged image of dash box. It shows clearly the existence of a bifurcation point in this coupled system with coupling voltage above the threshold.

In the above studies, the capacitance $C$ between sensing electrodes for detecting charge is assumed to be constant. However, due to the relative motion of the two sensing electrodes, capacitance $C$ may vibrate. The effect of vibration of $C$ on charge detection has been done analytically. Influence of vibration of $C$ is proved to be negligible (see in Supplement Material S2).

*The Method of Real-time Charge Detection*

A closed-loop circuit (Fig.3a) is setup for real-time charge detection. After amplification, filter and phase shift, the vibration signal of resonator R1 is transformed to a square signal to drive the resonator. Coupling voltage $V_C$ including polarization voltage $V_P$ and dynamic voltage signal $V_S$ is loaded on the body of resonator R2. Here, the bias voltage $V_P$ is set as 14 V to make sure that the couple voltage exceeds the threshold value. We plotted the step response for varying $V_C$ as shown in Fig.3b. $V_S$ is varied by a step of 0.005 V (52.3E-3 fC). It is also seen that the response frequency drift has high resolution for small varying $V_C$. The frequency fluctuation $\delta f$ equals to Allan deviation $\sigma_A$ multiply characteristic frequency $f_c$. According to the experimental data ($f_c \approx$237.88 KHz in Fig.3b and $\sigma_A$ =13.6 ppb in Fig.3c), the resolution in the closed-loop can be calculated as 2.051E-3 fC via $R = C\delta f/K_{sf}$, which is equivalent to the charge provided by about 13 electrons. The resolution of our charge detection has nearly 3 orders improvement compared to the counterparts reported in the earlier literatures[5,6,7].

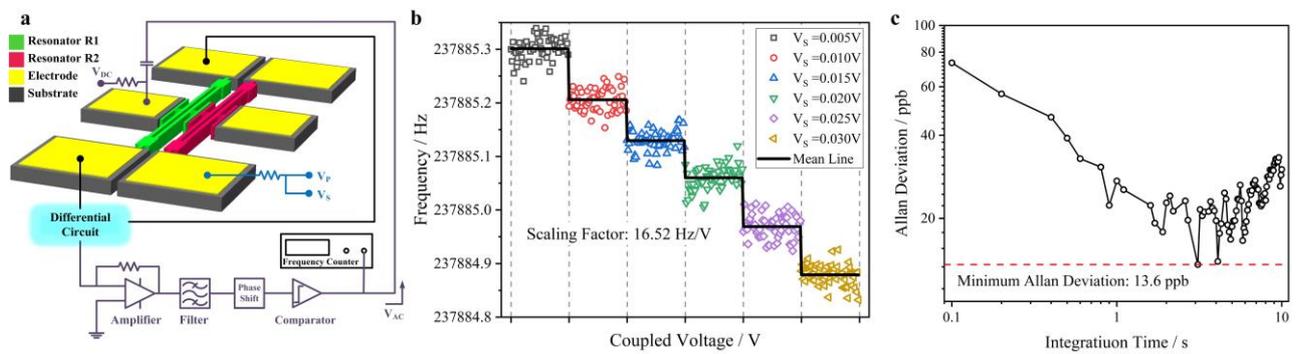

**Fig.3 Closed-loop experiment.** All tests are processed with the same conditions of $V_{AC} = 316.2 mV$ (driving voltage) and $V_{DC} = 18V$ (polarization voltage). **a** Schematic drawing of the closed-loop measurement circuit. **b** Detected step response with different coupling voltage step. The symbols represent experimental data of response frequency from frequency counter, and the solid line denotes the mean value. **c** Allan deviation of the micro-oscillator.

## *Extra Application for ULF Signal detection*

Weak and ULF signal transducer is widely used for vibratory analysis and malfunction inspection of an engineering structure[21], earthquake forecast[22], or geologic reconnoitering[23]. As an extra application, the coupled system can be used to track a weak and ULF signal. In this application,, $V_S$ is set as a dynamic sinusoidal signal. Dynamic responses are detected and plotted for weak and ULF dynamic signal $V_S$ with different amplitude and frequency, as shown in Fig. 4a. The left vertical axis represents frequency variation, and the right vertical axis represents corresponding voltage signal. Obviously, the weak and ULF signal can be accurately identified. The sampling rate $f_{SR}$ of the frequency counter is 10 Hz in this test. Theoretically, the circuit can detect a signal with $2f_S < f_{SR}$ considering the aliasing effect[24].

Furthermore, this experiment circuit is also used to detect mixed weak and ULF signal. Here, we set $V_S$ as a mixed two sinusoidal signals with different amplitude and frequency. Fig.4b plotted the frequency vibration and corresponding voltage output. Fig.4c plotted the corresponding frequency domain curves. The mixed dynamic signal is also accurately identified.

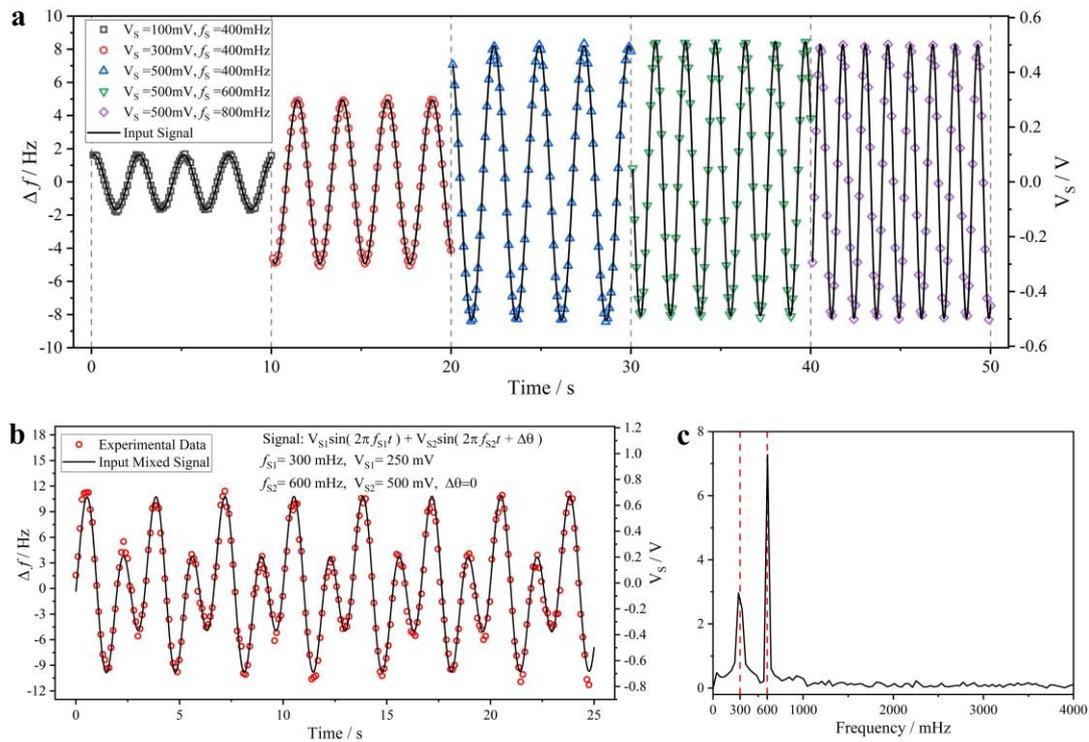

**Fig.4 Signal detection experiment. a** Detected dynamic response under several sine signals. The amplitude $V_S$ of sine signal is varied from 100 mV to 500 mV with 200 mV step length and the frequency $f_S$ of sine signal is varied from 400 mHz to 800 mHz with 200 mHz step length. **b** Detected dynamic response under a mixed signal. **c** Signal analysis through Fast Fourier Transformation.

## Material and Method

### Fabrication of the Micro-resonator

The micro-resonators were fabricated through a commercial silicon-on-insulator (SOI)–MUMPs foundry process from MEMSCAP. Detailed information about the processing technology can be found in Ref.25.

### Measurement scheme

All the measurement are tested in a vacuum chamber at a pressure below 3 Pa at room temperature. The experiment consists of two independent tests: open-loop test and closed-loop test. The motion of the resonator is detected through piezoresistive effect for both tests. Direct voltage $\pm V_D$ are applied on the anchor pads of the actuated resonator R1 to generate a current through the resonator beam. Meanwhile, $\pm V_D$ also make sure the electric potential of the electrode of resonator R1 equals zero, thus the coupling strength can be represented by $V_C$ directly. We controlled the coupling voltage $V_C$ through a SourceMeter (KEITHLEY 2400) accurately. In open-loop measurement circuit, a combination of $V_{AC}$ and $V_{DC}$ is loaded on one electrode to drive resonator R1 to vibration. The vibration can be sensed through a network analyzer (Agilent E5061B). Sustaining the beam's vibration at a fixed amplitude and a fixed frequency, a closed-loop circuit is setup for real-time detection. Due to the linear relation between resonant frequency and coupled voltage mentioned above, a weak dynamic signal can be detected by observing the changing oscillation frequency in the closed-loop circuit. The vibration signal of resonator R1 is extracted and magnified by differential amplifier. After amplification, the resulting signal is filtered and phase shifted. Then a square wave signal with a fixed amplitude from the comparator is applied on the electrode to electrostatically drive the resonator. A frequency counter (Agilent 53230A) is used to log the frequency output.

*Data analysis*

In open-loop experiment, we must plot complete resonant amplitude curve so as to find a peak frequency. Then, the variation of charge can be calculated through peak frequency drift. However, in closed-loop experiment, we can obtain the response frequency drift directly and realize real-time detection.

**Acknowledgements**

This work was financially supported by the National Natural Science Foundation of China (51575439, 11621062, 51421004, 11772293 and 11432012), National Key R & D Program of China (2018YFB2002303), Key research and development

program of Shaanxi Province (2018ZDCXL-GY-02-03) and 111project (B12016). We also appreciate the support from the Collaborative Innovation Center of High-End Manufacturing Equipment and the International Joint Laboratory for Micro/Nano Manufacturing and Measurement Technologies.

**Conflict of Interest**

The authors declare that they have no conflict of interest.